\documentclass[preprintnumbers,article,amsmath,amssymb,floatfix,14pt,prd,onecolumn,
superscriptaddress,nofootinbib]{revtex4-2}
\usepackage{bm}
\usepackage{amsfonts}
\usepackage{latexsym}
\usepackage[latin1]{inputenc}
\usepackage{graphicx}
\usepackage{amsmath}
\usepackage{palatino}
\usepackage{mathpazo}
\usepackage{textcomp}
\usepackage{float}
\usepackage{booktabs}
\usepackage{dcolumn}
\usepackage{ragged2e}
\usepackage{hyperref}
\hypersetup{colorlinks,citecolor=blue}
\hypersetup{colorlinks=true,linkcolor=red,filecolor=magenta,    urlcolor=blue}
\usepackage{amsmath}
\usepackage{xcolor}
\usepackage{orcidlink}
\usepackage{epsfig}
\usepackage{subfigure}
\usepackage{commath}

\def\jnl@style{\it}
\def\aaref@jnl#1{{\jnl@style#1}}

\def\aaref@jnl#1{{\jnl@style#1}}

\def\aj{\aaref@jnl{AJ}}                   
\def\apj{\aaref@jnl{ApJ}}                 
\def\apjl{\aaref@jnl{ApJ}}                
\def\apjs{\aaref@jnl{ApJS}}               
\def\apss{\aaref@jnl{Ap\&SS}}             
\def\aap{\aaref@jnl{A\&A}}                
\def\aapr{\aaref@jnl{A\&A~Rev.}}          
\def\aaps{\aaref@jnl{A\&AS}}              
\def\mnras{\aaref@jnl{Mon.~Not.~Roy.~Astron.~Soc.}}             
\def\prd{\aaref@jnl{Phys.~Rev.~D}}        
\def\prc{\aaref@jnl{Phys.~Rev.~C}}  
\def\prl{\aaref@jnl{Phys.~Rev.~Lett.}}    
\def\qjras{\aaref@jnl{QJRAS}}             
\def\skytel{\aaref@jnl{S\&T}}             
\def\ssr{\aaref@jnl{Space~Sci.~Rev.}}     
\def\zap{\aaref@jnl{ZAp}}                 
\def\nat{\aaref@jnl{Nature}}              
\def\aplett{\aaref@jnl{Astrophys.~Lett.}} 
\def\apspr{\aaref@jnl{Astrophys.~Space~Phys.~Res.}} 
\def\physrep{\aaref@jnl{Phys.~Rep.}}      
\def\physscr{\aaref@jnl{Phys.~Scr}}       
\def\commat{\aaref@jnl{Comm.~Math.~Phys.}}              
\def\science{\aaref@jnl{Science}}               
\def\cqg{\aaref@jnl{Classical Quant.~Grav.}}            
\def\jpcs{\aaref@jnl{JPCS}}                                     
\def\ijmpd{\aaref@jnl{Int.~J.~Mod.~Phys.~D}}                    
\def\grg{\aaref@jnl{Gen.~Relat.~Gravit.}}               
\def\rpp{\aaref@jnl{Rep.~Prog.~Phys.}}          
\def\npa{\aaref@jnl{Nucl.~Phys.~A}}        
\def\lrr{\aaref@jnl{Living Rev.~Rel.}}                   
\def\jcap{\aaref@jnl{J.~Cosmology Astropart.~Phys.}}    
\def\rmp{\aaref@jnl{Rev.~Mod.~Phys.}}   
\def\epjc{\aaref@jnl{Eur.~Phys.~J.~C}} 
\def\plb{\aaref@jnl{~Phy.~Lett.~B}} 
\def\mpla{\aaref@jnl{Mod.~Phy.~Lett.~A}} 
\def\arxiv{\aaref@jnl{arxiv.org}}

\allowdisplaybreaks[1]

\begin{document}
\color{black}       
\title{Dynamical system analysis of scalar field cosmology in coincident $f(Q)$ gravity}

\author{Sayantan Ghosh\orcidlink{0000-0002-3875-0849}}
\email{sayantanghosh.000@gmail.com}
\affiliation{Department of Mathematics, Birla Institute of Technology and Science-Pilani,\\ Hyderabad Campus, Hyderabad-500078, India.}

\author{Raja Solanki\orcidlink{0000-0001-8849-7688}}
\email{rajasolanki8268@gmail.com}
\affiliation{Department of Mathematics, Birla Institute of Technology and
Science-Pilani,\\ Hyderabad Campus, Hyderabad-500078, India.}

\author{P.K. Sahoo\orcidlink{0000-0003-2130-8832}}
\email{pksahoo@hyderabad.bits-pilani.ac.in}
\affiliation{Department of Mathematics, Birla Institute of Technology and Science-Pilani,\\ Hyderabad Campus, Hyderabad-500078, India.}

%
\date{\today}
\begin{abstract}
In this article, we investigate scalar field cosmology in the coincident $f(Q)$ gravity formalism. We calculate the motion equations of $f(Q)$ gravity under the flat Friedmann-Lema\^{i}tre-Robertson-Walker background in the presence of a scalar field. We consider a non-linear $f(Q)$ model, particularly $f(Q)=-Q+\alpha Q^n$, which is nothing but a polynomial correction to the STEGR case. Further, we assumed two well-known specific forms of the potential function, specifically the exponential from $V(\phi)= V_0 e^{-\beta \phi}$ and the power-law form $V(\phi)= V_0\phi^{-k}$. We employ some phase-space variables and transform the cosmological field equations into an autonomous system. We calculate the critical points of the corresponding autonomous systems and examine their stability behaviors. We discuss the physical significance corresponding to the exponential case for parameter values $n=2$ and $n=-1$ with $\beta=1$, and $n=-1$ with $\beta=\sqrt{3}$. Moreover, we discuss the same corresponding to the power-law case for the parameter value $n=-2$ and $k=0.16$. We also analyze the behavior of corresponding cosmological parameters such as scalar field and dark energy density, deceleration, and the effective equation of state parameter. Corresponding to the exponential case, we find that the results obtained for the parameter constraints in Case III is better among all three cases, and that represents the evolution of the universe from a decelerated stiff era to an accelerated de-Sitter era via matter-dominated epoch. Further, in the power-law case, we find that all trajectories exhibit identical behavior, representing the evolution of the universe from a decelerated stiff era to an accelerated de-Sitter era. Lastly, we conclude that the exponential case shows better evolution as compared to the power-law case.
\end{abstract} 

\maketitle


\section{Introduction}\label{sec1}
\justifying

Modified gravity is a prominent route to enhance our understanding of the evolution of the Universe, incorporating its early stages (inflation) and later phases (dark energy) marked by accelerated expansion \cite{CANT}. Additionally, it may address potential observational conflicts \cite{COSI}. These theories involve constructing extensions and modifications of the General Relativity that introduce additional degrees of freedom. These modifications can potentially provide corrections at both the cosmological background and perturbation levels, offering a more comprehensive description of the Universe's behavior. There exist various methods for constructing such gravitational modifications. One different category of gravitational modifications emerges when utilizing the equivalent formulation of gravity that incorporates non-metricity, which is widely known as symmetric teleparallel gravity \cite{NEST}. This approach employs a generic affine connection with vanishing torsion and vanishing curvature with respect to Levi-Civita connection, while relaxing the metricity condition with respect to the generic connection. This formulation has recently been extended to $f(Q)$ gravity \cite{JIM-1}. This extension of the symmetric teleparallel formalism has gained attention from the cosmology community as a potential avenue for exploring new physics beyond the conventional $\Lambda$CDM cosmology. Particular forms of the $f(Q)$ function have been demonstrated to alleviate the $\sigma 8$ tension \cite{BARR}, whereas others enable a more accurate representation of cosmological data \cite{ANAG,ARORA,NUNES}. Recently, some interesting cosmological implications of the $f(Q)$ gravity in the different context have been appeared, for instance, Black hole physics \cite{RODR,LAVI-1}, Neutrino physics \cite{NEOM}, Quantum cosmology \cite{CAPE-1,PALIA}, Bouncing Cosmology \cite{GADB}, Inflation \cite{CAPE-2}, Phantom cosmology \cite{ANDER}, Astrophysical objects \cite{SNEHA,Zinnat}, Cosmological perturbations \cite{ET}, BBN constraints \cite{ANAG-2}, and many others \cite{DE-1,DE-2,PALIA-2,HOH,WOM-2}.\\

Scalar fields play a significant role in describing the physical properties of the universe, particularly in the context of the inflationary scenario \cite{ALAN}, and they can offer explanations for the cosmic late-time acceleration. While the $\Lambda$CDM model is highly consistent with observational data, successfully describing structure formation, it has yet to quantify quantum vacuum fluctuations effectively \cite{ZEL,WENB}. This is the key inspiration for proposing the dark energy candidate as an alternative to $\Lambda$. Various examples include the quintessence field \cite{quint1,quint2}, a quintom scalar field \cite{quintom1,quintom2}, a phantom scalar field \cite{phant}, and multi-scalar field models \cite{mult}. If one assumes the cosmological constant $\Lambda$ is coming from a single (or multiple) scalar field, then one can solve the cosmological constant problem, as the scalar field can decay the vacuum energy to stop the exponential increase. However, there exists a tension between the values of the Hubble constant measured from early observations (such as Planck \cite{Planck}) and estimated via local observations (such as SH0ES \cite{Riess}). One potential solution to this tension involves assuming extensions beyond the $\Lambda$CDM. This article explores scalar field cosmology in the modified symmetric teleparallel gravity background, utilizing the dynamical system approach. The modified $f(Q)$ function is now responsible for the observed accelerating scenario, whereas the quantum field theory prediction of the tiny value of $\Lambda$ ($\sim 10^{-120}$) cannot be described by such gravitational modification and thus the addition of a scalar field take into account for quintessence scenario.\\

It is widely recognized that when investigating cosmological models, introducing auxiliary variables allows for transforming cosmological equations into an autonomous dynamical system \cite{COPE}. This results in a system $X' =f(X)$, where $X$ represents the column vector consisting of auxiliary variables and $f(X)$ denotes the vector field. The stability analysis of the given autonomous involves several steps. Initially, critical points (or equilibrium points) $X_c$ are identified by setting $X'=0$.  Then, we consider linear perturbations around the critical point $X_c$ as $X=X_c+P$, where $P$ represents the column vector of perturbed auxiliary variables. As a result, one can have (up to first order) $P'=AP$, where $A$ is the matrix consisting of coefficients of the perturbed equations. Finally, the eigenvalues of the coefficient matrix $A$ determine the stability behavior of each hyperbolic critical point. A critical point $X_c$ is considered stable (unstable), or a saddle if the real parts of the corresponding eigenvalues are negative (positive), or have real parts with different signs. Several interesting outcomes in the context of modified gravity utilizing dynamical system method can be found in references \cite{DE-3,WOM,Mishra-1,Mishra-2,HAMID}. The present article is organized as follows. In Sec \ref{sec2}, we present the mathematical background of extended symmetric teleparallel gravity with the Lagrangian density of a scalar field. In Sec \ref{sec3}, we present governing equations of motion under the FLRW background in the presence of a scalar field. Further in Sec \ref{sec4}, we perform a detailed dynamical analysis of two well-known scalar field potentials along with a non-linear $f(Q)$ cosmological background model. Finally in Sec \ref{sec5}, we highlight our findings.

\section{$f(Q)$ gravity formulation with scalar field}\label{sec2}
\justifying

We begin with the fact that a connection plays a vital role in the transportation of tensors across a manifold. In the realm of general relativity, which is built upon Riemannian geometry, the gravitational interactions are ruled by a symmetric connection referred to as the Levi-Civita connection. Nevertheless, a more generic connection comprises two components: an antisymmetric part and another component exhibiting the non-metricity condition. This extended affine connection can be expressed as follows \cite{TRIN}:

\begin{equation}\label{2a}
\Upsilon^\alpha_{\ \mu\nu}=\Gamma^\alpha_{\ \mu\nu}+K^\alpha_{\ \mu\nu}+L^\alpha_{\ \mu\nu},
\end{equation}
where the first term denotes the metric compatible Levi-Civita connection,
\begin{equation}\label{2b}
\Gamma^\alpha_{\ \mu\nu}\equiv\frac{1}{2}g^{\alpha\lambda}(g_{\mu\lambda,\nu}+g_{\lambda\nu,\mu}-g_{\mu\nu,\lambda})
\end{equation}
The second one is the contortion tensor representing the antisymmetric part of the affine connection, and that can estimate in terms of the torsion tensor $ T^\alpha_{\ \mu\nu}\equiv \Upsilon^\alpha_{\ \mu\nu}-\Upsilon^\alpha_{\ \nu\mu}$ as follows,
\begin{equation}\label{2c}
K^\alpha_{\ \mu\nu}\equiv\frac{1}{2}(T^{\alpha}_{\ \mu\nu}+T_{\mu \ \nu}^{\ \alpha}+T_{\nu \ \mu}^{\ \alpha})
\end{equation}
, and the last one is the distortion tensor,
\begin{equation}\label{2d}
L^\alpha_{\ \mu\nu}\equiv\frac{1}{2}(Q^{\alpha}_{\ \mu\nu}-Q_{\mu \ \nu}^{\ \alpha}-Q_{\nu \ \mu}^{\ \alpha})
\end{equation}	
which is expressed in terms of non-metricity tensor, 
\begin{equation}\label{2e}
Q_{\alpha\mu\nu}\equiv\nabla_\alpha g_{\mu\nu} = \partial_\alpha g_{\mu\nu}-\Upsilon^\beta_{\,\,\,\alpha \mu}g_{\beta \nu}-\Upsilon^\beta_{\,\,\,\alpha \nu}g_{\mu \beta}
\end{equation} 
In addition, we define the superpotential tensor as
\begin{equation}\label{2f}
4P^\lambda\:_{\mu\nu} = -Q^\lambda\:_{\mu\nu} + 2Q_{(\mu}\:^\lambda\:_{\nu)} + (Q^\lambda - \tilde{Q}^\lambda) g_{\mu\nu} - \delta^\lambda_{(\mu}Q_{\nu)}.
\end{equation}
where $Q_\alpha = Q_\alpha\:^\mu\:_\mu $ and $ \tilde{Q}_\alpha = Q^\mu\:_{\alpha\mu} $ are non-metricity vectors.
Now by contracting the superpotential tensor with the non-metricity tensor, we obtain non-metricity scalar in a more confine form,
\begin{equation}\label{2g}
Q = -Q_{\lambda\mu\nu}P^{\lambda\mu\nu}. 
\end{equation}
We know that the curvature tensor can be estimated as
\begin{equation}\label{2h}
R^\alpha_{\: \beta\mu\nu} = 2\partial_{[\mu} \Upsilon^\alpha_{\: \nu]\beta} + 2\Upsilon^\alpha_{\: [\mu \mid \lambda \mid}\Upsilon^\lambda_{\nu]\beta}
\end{equation} 
Now by using the affine connection \eqref{2a}, one can have
\begin{equation}\label{2i}
R^\alpha_{\: \beta\mu\nu} = \mathring{R}^\alpha_{\: \beta\mu\nu} + \mathring{\nabla}_\mu X^\alpha_{\: \nu \beta} - \mathring{\nabla}_\nu X^\alpha_{\: \mu \beta} + X^\alpha_{\: \mu\rho} X^\rho_{\: \nu\beta} - X^\alpha_{\: \nu \rho} X^\rho_{\: \mu\beta}
\end{equation}
Here $\mathring{R}^\alpha_{\: \beta\mu\nu}$ and $\mathring{\nabla}$ are described in terms of the Levi-Civita connection \eqref{2b}, and $X^\alpha_{\ \mu\nu}=K^\alpha_{\ \mu\nu}+L^\alpha_{\ \mu\nu}$.
On applying suitable contractions on curvature term and torsion free constraint $ T^\alpha_{\ \mu\nu}=0$ in the equation \eqref{2i}, we have
\begin{equation}\label{2j}
R=\mathring{R}-Q + \mathring{\nabla}_\alpha \left(Q^\alpha-\tilde{Q}^\alpha \right)   
\end{equation}
where $\mathring{R}$ is the usual Ricci scalar evaluated in terms of the Levi-Civita connection.
Further on employing teleparallel constraint $R=0$, we acquire curvature free teleparallel geometries, and hence relation \eqref{2j} becomes
\begin{equation}\label{2k}
\mathring{R}=Q - \mathring{\nabla}_\alpha \left(Q^\alpha-\tilde{Q}^\alpha \right)   
\end{equation}
The relation obtained in the equation \eqref{2k} indicates that the Ricci scalar curvature differs from the non-metricity scalar by a boundary term. The equation \eqref{2k} reveals that the gravity theory incorporates only non-metricity scalar $Q$ in the action differs from the Einstein's GR by a boundary term. This indicates that STEGR presents an equivalent formulation to GR, and hence the theory is known as symmetric teleparallel equivalent to GR \cite{KUHN}. 

Now, we present the action for $f(Q)$ gravity which is a generalization to STEGR theory, in presence of a scalar field, 
\begin{equation}\label{2l}
\mathcal{S}=\int\frac{1}{2}\,f(Q)\sqrt{-g}\,d^4x+\int \mathcal{L}_{\phi}\,\sqrt{-g}\,d^4x\, ,
\end{equation}
where $g=\text{det}(g_{\mu\nu})$, $f(Q)$ is an arbitrary function of the non-metricity scalar $Q$, and $\mathcal{L}_{\phi}$ is the Lagrangian density of a scalar field $\phi$ given by \cite{BAHA},

\begin{equation}\label{2m}
\mathcal{L}_{\phi} = -\frac{1}{2} g^{\mu \nu} \partial_\mu \phi  \partial_\nu \phi -V(\phi)
\end{equation}
Here $V(\phi)$ represents a potential for the field $\phi$. We obtain the following governing field equation by varying the action \eqref{2l} with respect to the metric,

\begin{equation}\label{2n}
\frac{2}{\sqrt{-g}}\nabla_\lambda (\sqrt{-g}f_Q P^\lambda\:_{\mu\nu}) + \frac{1}{2}g_{\mu\nu}f+f_Q(P_{\mu\lambda\beta}Q_\nu\:^{\lambda\beta} - 2Q_{\lambda\beta\mu}P^{\lambda\beta}\:_\nu) = -T_{\mu\nu}^{\phi}
\end{equation}

where $f_Q=\frac{df}{dQ}$ and $T_{\mu\nu}^{\phi}$ represents the stress-energy tensor of the scalar field given as
\begin{equation}\label{2o}
 T_{\mu\nu}^{\phi}= \partial_\mu \phi  \partial_\nu \phi -\frac{1}{2} g_{\mu \nu} g_{\alpha \beta} \partial^\alpha \phi  \partial^\beta \phi -  g_{\mu \nu} V(\phi)
\end{equation}
Moreover, we obtain the following equation of motion for the scalar field i.e. Klein-Gordon equation from the Euler-Lagrangian equation for the Lagrangian density given by \eqref{2m}
\begin{equation}\label{2p}
\square \phi - V,_\phi =0
\end{equation}
Here $\square$ denotes the d'Alembertian and $V,_\phi = \frac{\partial V}{\partial \phi}$. Further, on varying the action \eqref{2l} with respect to the connection (similar to Palatini prescription), we have, 
\begin{equation}\label{2q}
\nabla_\mu \nabla_\nu (\sqrt{-g}f_Q P^{\mu\nu}\:_\lambda) =  0 
\end{equation}

\section{Equations of Motion}\label{sec3}
\justifying
 
We begin with the following flat FLRW line element in order to probe the cosmological implications under the assumption of spatial isotropy and homogeneity of the universe,
\begin{equation}\label{3a}
ds^2= -dt^2 + a^2(t)[dx^2+dy^2+dz^2]    
\end{equation}
Here $a(t)$ is a measure of expansion of the universe. Beginning with the teleparallel constraint that corresponds to a flat geometry characterizing a pure inertial connection, one can execute a gauge transformation parameterized by $\Lambda^\alpha_\mu$  \cite{JIM-2},
\begin{equation}\label{3b}
 \Upsilon^\alpha_{\: \mu \nu}  = (\Lambda^{-1})^\alpha_{\:\: \beta} \partial_{[ \mu}\Lambda^\beta_{\: \: \nu ]}
\end{equation}
Consequently, one can express the generic affine connection in the following manner, by utilizing the general element of $ GL(4,\mathbb{R}) $ characterized by the transformation $ \Lambda^\alpha_{\: \: \mu}=\partial_\mu \zeta^\alpha$, where $ \zeta^\alpha $ is an arbitrary vector field,
\begin{equation}\label{3c}
\Upsilon^\alpha_{\: \mu \nu} = \frac{\partial x^\alpha}{\partial \zeta^\rho} \partial_\mu \partial_\nu \zeta^\rho
\end{equation}
This reveals the possibility of eliminating the connection through a coordinate transformation. The coordinate transformation is responsible for eliminating the connection \eqref{3c} is termed gauge coincident. We utilize the coincident gauge in the present manuscript. Hence, the non-metricity scalar corresponds to the metric \eqref{3a} becomes $Q=6H^2$.\\
The stress-energy tensor for the perfect fluid distribution reads as
\begin{equation}\label{3d}
T_{\mu\nu}=(\rho+p)u_\mu u_\nu + pg_{\mu\nu}
\end{equation}
where $u^\mu=(1,0,0,0)$ are components of the four velocities. On comparing equation \eqref{3d} and \eqref{2o}, we have,
\begin{equation}\label{3e}
\rho=-\frac{1}{2}g_{\alpha \beta}\partial^\alpha \phi \partial^\beta \phi +V(\phi)
\end{equation}
\begin{equation}\label{3f}
p=-\frac{1}{2}g_{\alpha \beta}\partial^\alpha \phi \partial^\beta \phi -V(\phi)
\end{equation}
As the scalar field considered here does not depend on the spatial coordinates, we have following expressions for the energy density and pressure component of the scalar field,
\begin{equation}\label{3g}
\rho_{\phi}=\frac{1}{2}\dot{\phi}^2+V(\phi)
\end{equation}
\begin{equation}\label{3h}
p_{\phi}=\frac{1}{2}\dot{\phi}^2-V(\phi)
\end{equation}
and the corresponding equation of state parameter can be written as,
\begin{equation}\label{3i}
\omega_{\phi}=\frac{p_{\phi}}{\rho_{\phi}}=\frac{\frac{1}{2}\dot{\phi}^2-V(\phi)}{\frac{1}{2}\dot{\phi}^2+V(\phi)}
\end{equation}
Moreover, corresponding to the metric \eqref{3a}, the Klein-Gordon equation \eqref{2p} becomes,
\begin{equation}\label{3j}
\ddot{\phi}+3H\dot{\phi}+V_{,\phi}=0
\end{equation}
We obtain the following Friedmann like equations governing the gravitational interactions under the $f(Q)$ gravity background in the presence of scalar field,
\begin{equation}\label{3k}
3H^2=\frac{1}{2f_Q} \left( -\rho_{\phi}+\frac{f}{2}  \right)
\end{equation}
\begin{equation}\label{3l}
    \dot{H}+3H^2+ \frac{\dot{f_Q}}{f_Q}H = \frac{1}{2f_Q} \left( p_{\phi}+\frac{f}{2} \right)
\end{equation}
For the $f(Q)$ functional $f(Q)=-Q+\Psi(Q)$, we can rewrite the Friedmann equations \eqref{3k}-\eqref{3l} as (where we can recover ordinary GR by putting $\Psi=0$)
\begin{equation}\label{3m}
    3H^2= \rho_\phi + \rho_{de}
\end{equation}
\begin{equation}\label{3n}
    \dot{H}=-\frac{1}{2} [\rho_\phi + p_\phi+\rho_{de}+p_{de}]
\end{equation}
where $\rho_{de}$ and $p_{de}$ represents the energy density and pressure of the dark energy component evolving due to geometry of spacetime,
\begin{equation}\label{3o}
    \rho_{de}=-\frac{\Psi}{2}+ Q\Psi_Q
\end{equation}
\begin{equation}\label{3p}
    p_{de}=-\rho_{de}-2\dot{H} \left( \Psi_Q+2Q\Psi_{QQ} \right)
\end{equation}

\section{The Cosmological Model and Dynamical System Analysis}\label{sec4}
\justifying

In this section, we attempt to examine some specific class of scalar field potentials and $f(Q)$ functional forms, incorporating the phase-space techniques. In order to do so, we transform motion equations of considered cosmological settings, discussed in the previous section, into an autonomous system with the help of phase-space variables. We define the following phase-space variables,
\begin{equation}\label{4a}
x^2=\frac{\dot{\phi}^2}{6H^2}, \:\: y^2=\frac{V}{3H^2}, \:\: \text{and} \:\: s^2=\Omega_{de}=\frac{\rho_{de}}{3H^2}
\end{equation}
Now using the fact that $\Omega_\phi=\frac{\rho_\phi}{3H^2}=\frac{\dot{\phi}^2}{6H^2}+\frac{V}{3H^2}=x^2+y^2$ and the equation \eqref{3m}, we have following constraint,
\begin{equation}\label{4b}
x^2+y^2+s^2=1 
\end{equation}
Hence it follows that $\Omega_{de}=s^2=1-x^2-y^2=1-\Omega_\phi$ and the constraint $0\leq x^2+y^2 \leq 1$.\\

In order to obtain the close form of required autonomous system with general potential, we define another phase-space variable as follows,
\begin{equation}\label{4c}
\lambda=-\frac{V_{,\phi}}{V}    
\end{equation}
We obtain the following dynamical system  with respect to e-folding time $N=ln(a)$,
\begin{equation}\label{4d}
x'=-3x-x \frac{\dot{H}}{H^2} +\sqrt{\frac{3}{2}} \lambda y^2   
\end{equation}
\begin{equation}\label{4e}
y'=\frac{1}{2y} \left[ \frac{\dot{V}}{3H^2} - y^2 \frac{2\dot{H}}{H^2} \right]
\end{equation}
\begin{equation}\label{4f}
\lambda'=-\sqrt{6}\lambda^2 x (\Gamma-1)
\end{equation}
where $\Gamma=\frac{VV_{,\phi \phi}}{V_{,\phi}^2}$. Here using the equation \eqref{3n}, we have
\begin{equation}\label{4g}
\frac{\dot{H}}{H^2}=\frac{3x^2}{\left( \Psi_Q+2Q\Psi_{QQ}-1 \right)}
\end{equation}
Note that the dynamical system \eqref{4d}-\eqref{4f} still not in its closed form, as it depends on the choice of $f(Q)$ functional form. Further, it also appears that we gain nothing from this extra phase-space variable $\lambda$, due to the presence of a quantity $\Gamma$ in the given dynamical system, that directly relies on the scalar field $\phi$. Nevertheless, considering that both $\lambda$ and $\Gamma$ are functions dependent on the scalar field $\phi$, there exists an explicit relation between them i.e. if the function $\lambda(\phi)$ is invertible, allowing us to derive $\phi(\lambda)$, and thus we have $\Gamma$ as a function of $\lambda$.\\

We consider a power-law functional form $\Psi(Q)=\alpha Q^n$ i.e. $f(Q)=-Q+\alpha Q^n$, where $\alpha$ and $n$ are free parameters. The choice of considered functional form is appropriate in order to close the obtained dynamical system. Now, for the assumed $f(Q)$ function, the expression \eqref{4g} becomes
\begin{equation}\label{4h}
    \frac{\dot{H}}{H^2}=\frac{3x^2}{n \left( 1-x^2-y^2 \right) -1}
\end{equation}
Moreover, we have following expressions for two cosmological parameters that plays a crucial role in characterizing the evolutionary phase of expansion of the universe, namely deceleration and equation of state parameter,
\begin{equation}\label{4i}
q=-1- \frac{3x^2}{n \left( 1-x^2-y^2 \right) -1}
\end{equation}
\begin{equation}\label{4j}
\omega_{total}=-1- \frac{2x^2}{n \left( 1-x^2-y^2 \right) -1}
\end{equation}
Now in order to probe the cosmological implications of the considered scenario we choose two specific forms of potential function, namely exponential and power-law potentials, which are widely discussed in the literature in the different cosmological contexts.

\subsection{Exponential potential}
\justifying
We assume the following form of exponential potential,
\begin{equation}\label{s1}
V(\phi)= V_0 e^{-\beta \phi}
\end{equation}
The assumed exponential potential represents the most basic illustration of quintessence scenarios \cite{LAS, TBR}, as it is a slowly varying scalar field and could decay the cosmological constant $\Lambda$. Hence, it can be a suitable candidate to address the cosmological constant problem. Such potential arises naturally from the string theory compactifications discussed in \cite{DBL}. Also, this type of potential naturally arises from the Kaluza-Klein type compactifications, and their effect on anisotropic product cosmology has been investigated in the reference \cite{JYJ}. We note that the dynamical system analysis of scale-invariant solutions of the exponential scalar field coupled with a barotropic fluid system has been studied and has been shown that such a system can have stable scale-invariant stable fixed point \cite{COPE}. Now, by using equation \eqref{s1}, the dynamical variable \eqref{4c} becomes 
\begin{equation}\label{s2}
\lambda=\beta \:\: \text{with} \:\: \Gamma=1 
\end{equation}
,and hence the system \eqref{4d}-\eqref{4f} reduces to the following autonomous system,
\begin{equation}\label{s3}
x'=-3x \left[ 1+ \frac{x^2}{n \left( 1-x^2-y^2 \right)-1}  \right] + \sqrt{\frac{3}{2}} \beta y^2
\end{equation}
\begin{equation}\label{s4}
y'= -xy \left[ \sqrt{\frac{3}{2}} \beta + \frac{3x}{n \left( 1-x^2-y^2 \right)-1}  \right]
\end{equation}
The critical points and their physical behavior, corresponds to autonomous system \eqref{s3}-\eqref{s4} are presented in the Table \eqref{Table-1}.

\begin{figure}[h]
{\includegraphics[scale=0.52]{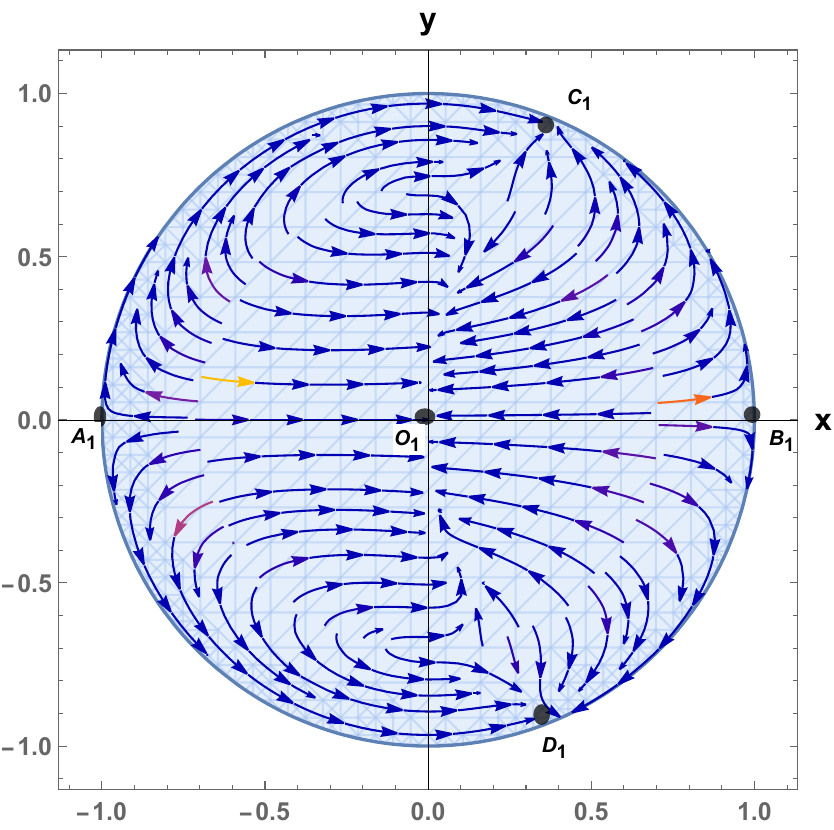}}
{\includegraphics[scale=0.52]{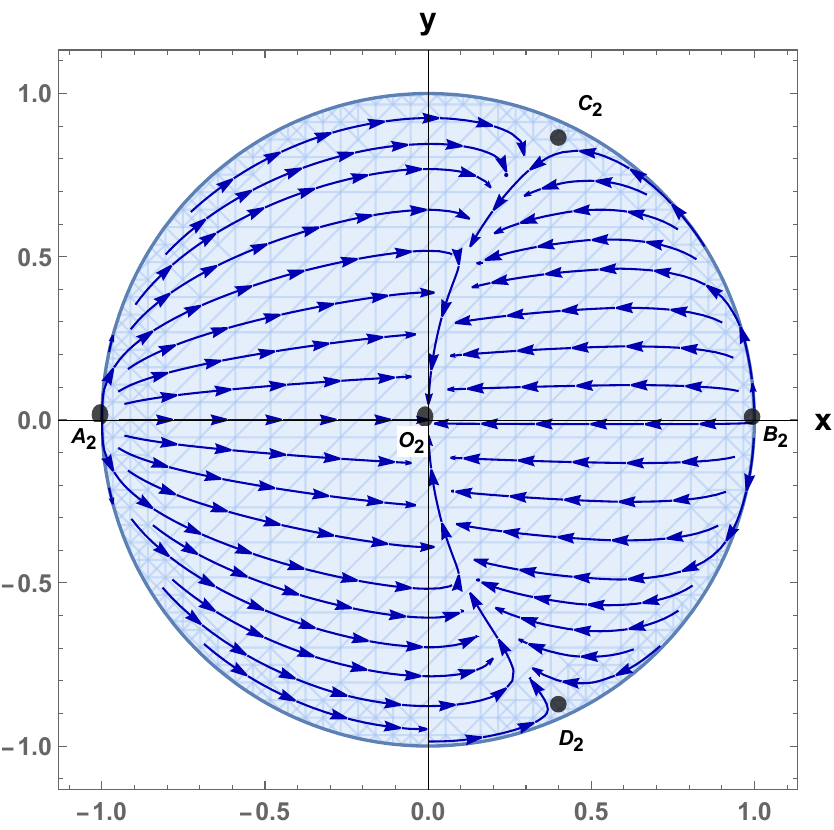}}
{\includegraphics[scale=0.52]{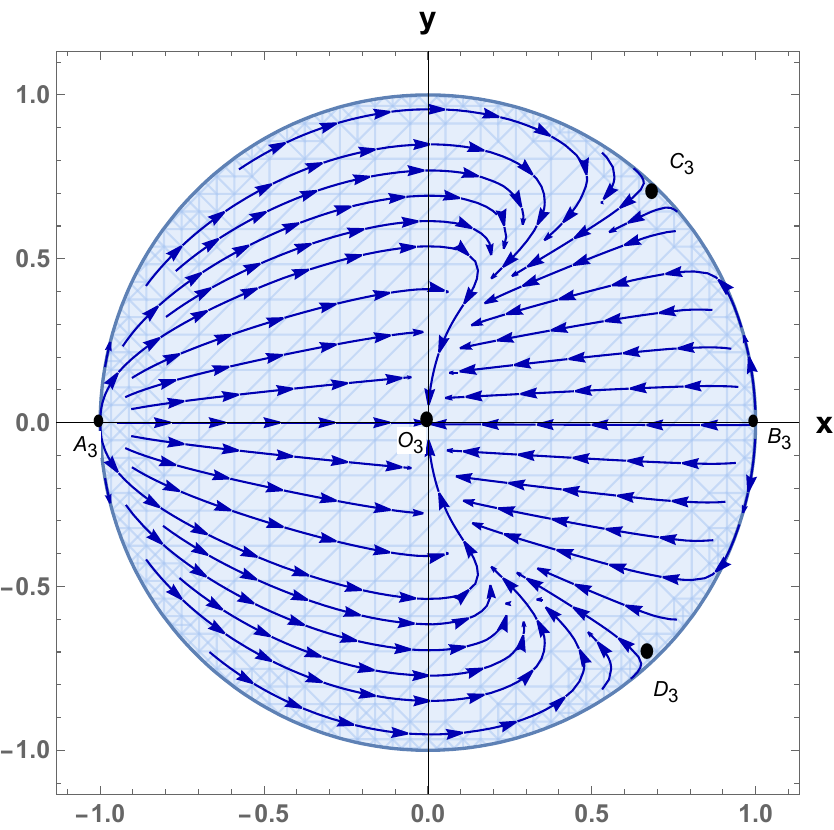}}
\caption{Phase-space plots for the case $n=2$ (left panel above) and $n=-1$ (right panel above), with $\beta=1$ , and for the case $n=-1$ with $\beta=\sqrt{3}$ (below) corresponding to the exponential potential.}\label{f1}
\end{figure}

\begin{figure}[H]
{\includegraphics[scale=0.34]{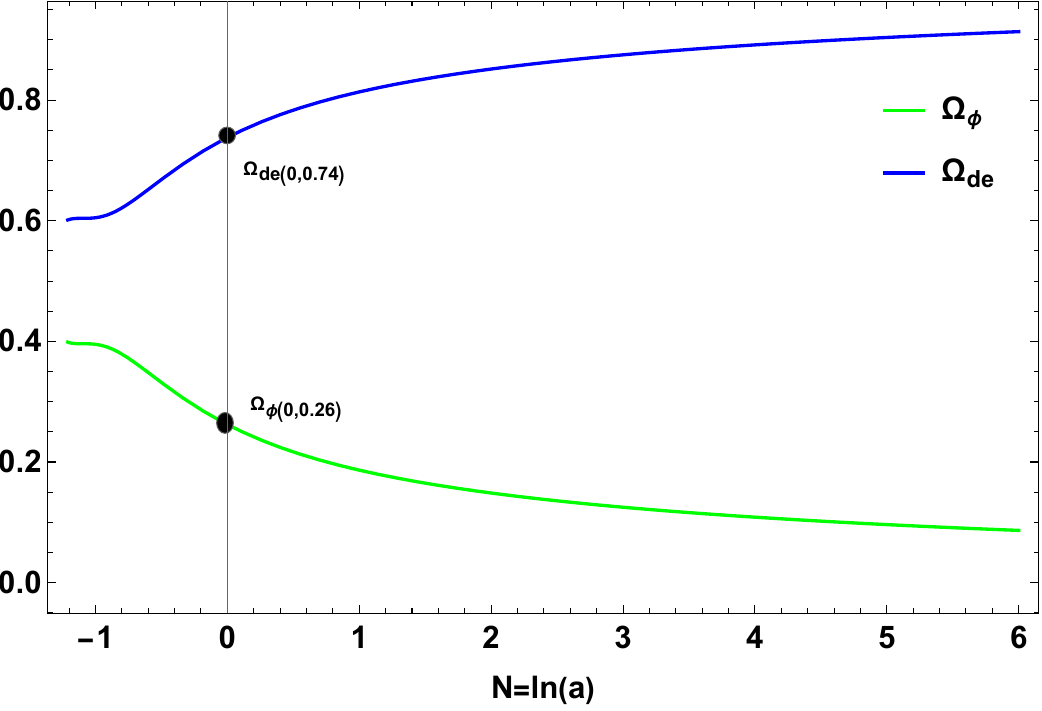}}
{\includegraphics[scale=0.34]{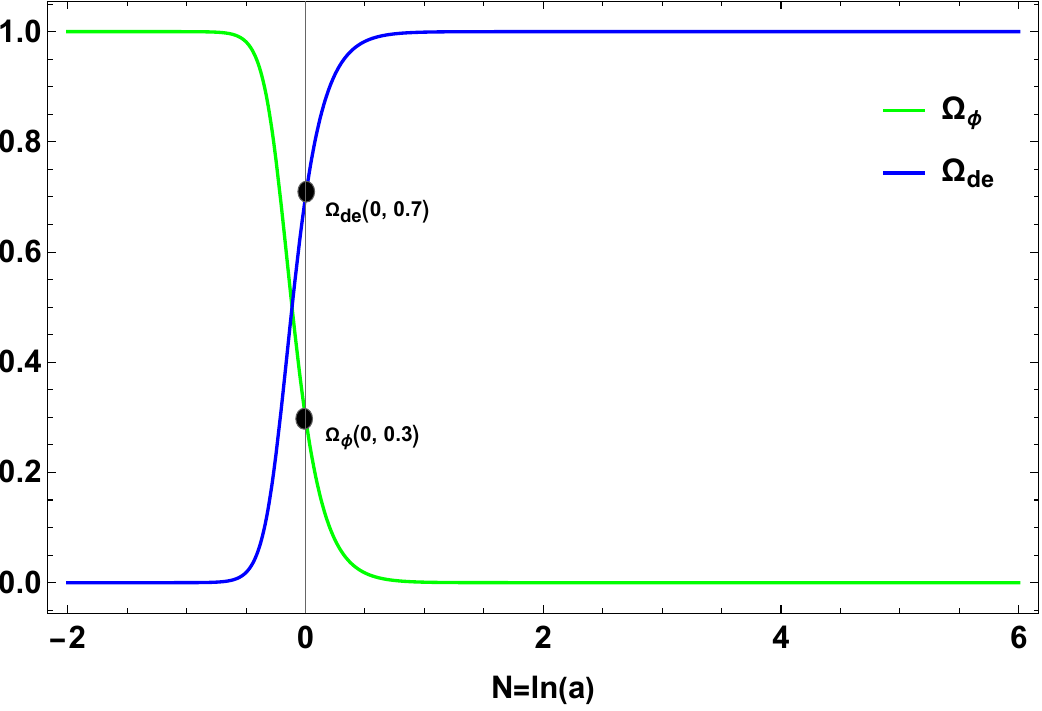}}
{\includegraphics[scale=0.34]{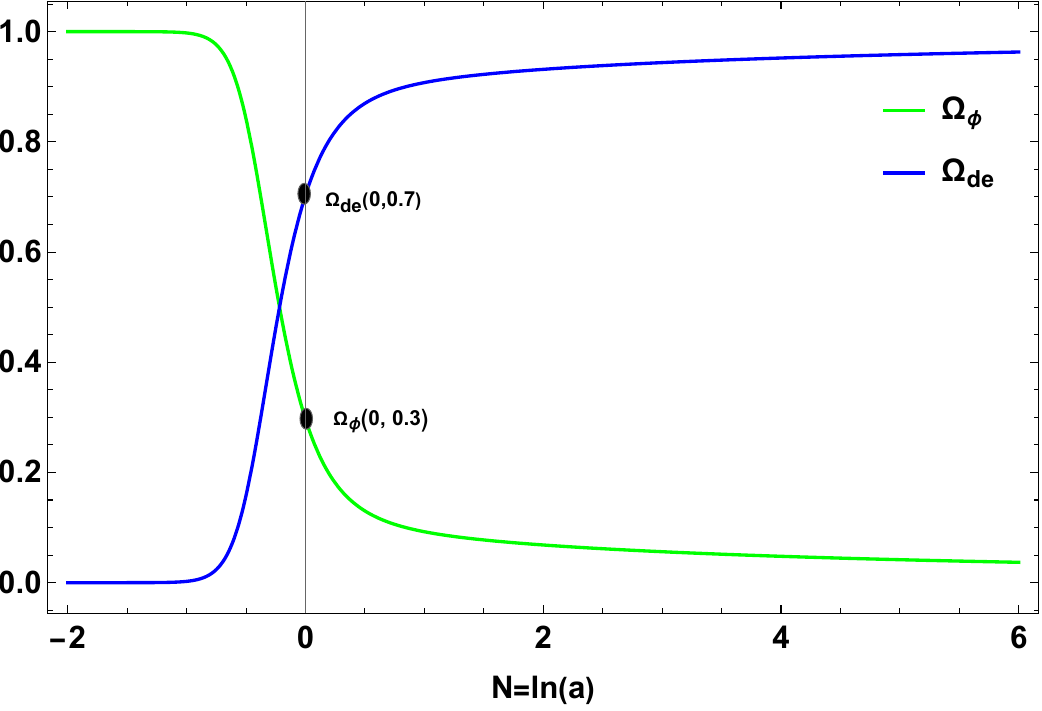}}
\caption{Evolutionary profile of the scalar field density and dark energy density for the case $n=2$ (left panel) and $n=-1$ (middle panel), with $\beta=1$, and for the case $n=-1$ with $\beta=\sqrt{3}$ (right panel) corresponding to the exponential potential.}\label{f2}
\end{figure}

\begin{figure}[H]
{\includegraphics[scale=0.34]{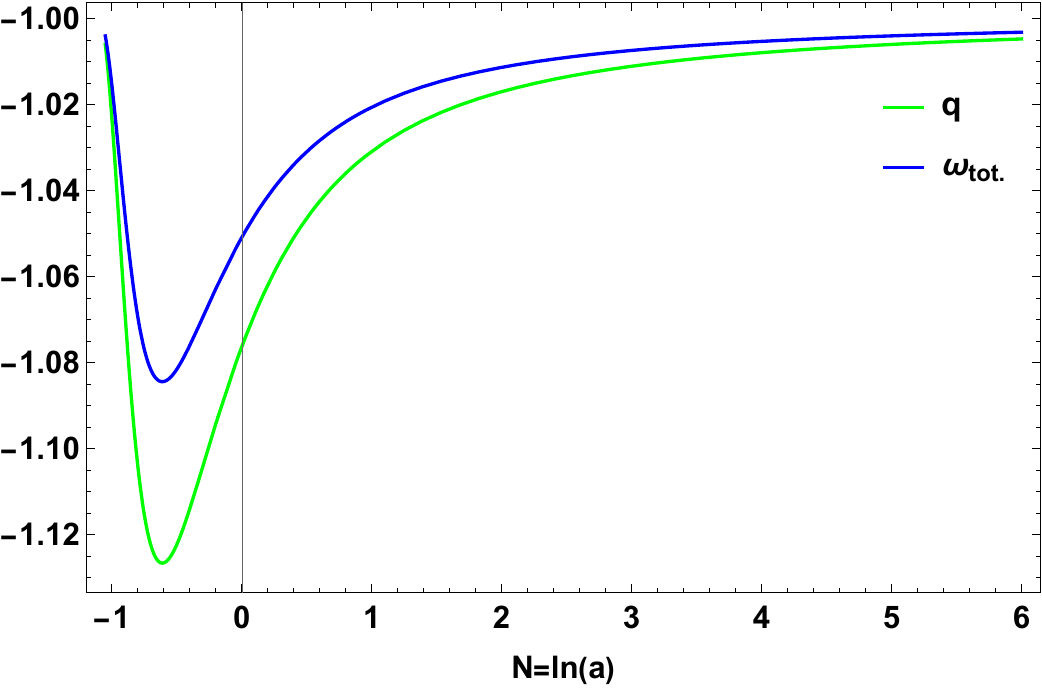}}
{\includegraphics[scale=0.34]{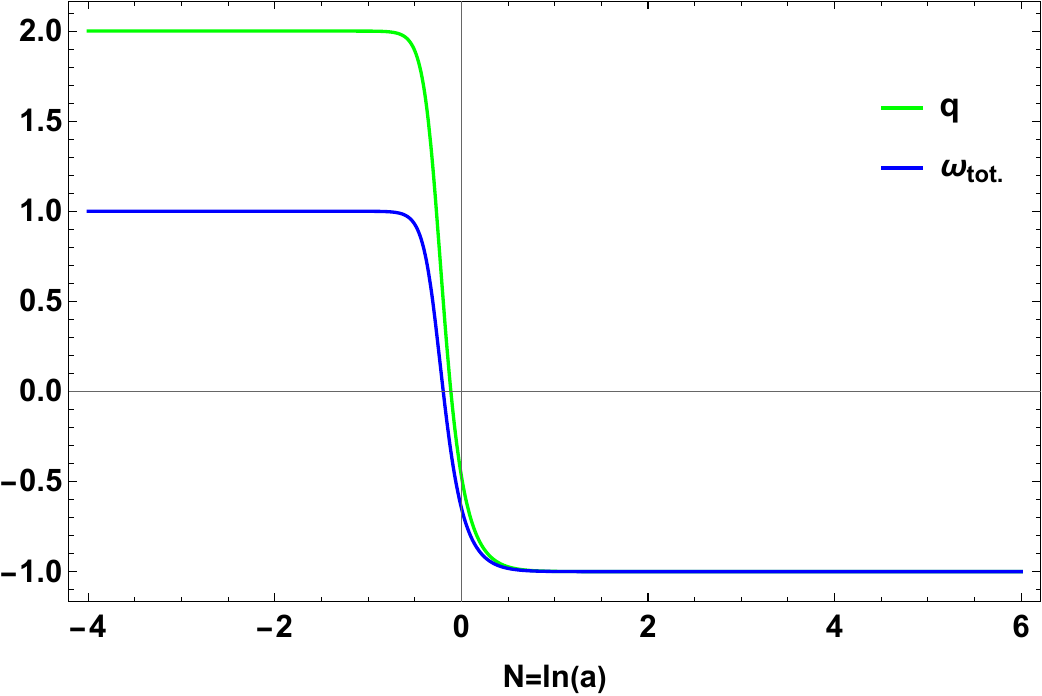}}
{\includegraphics[scale=0.34]{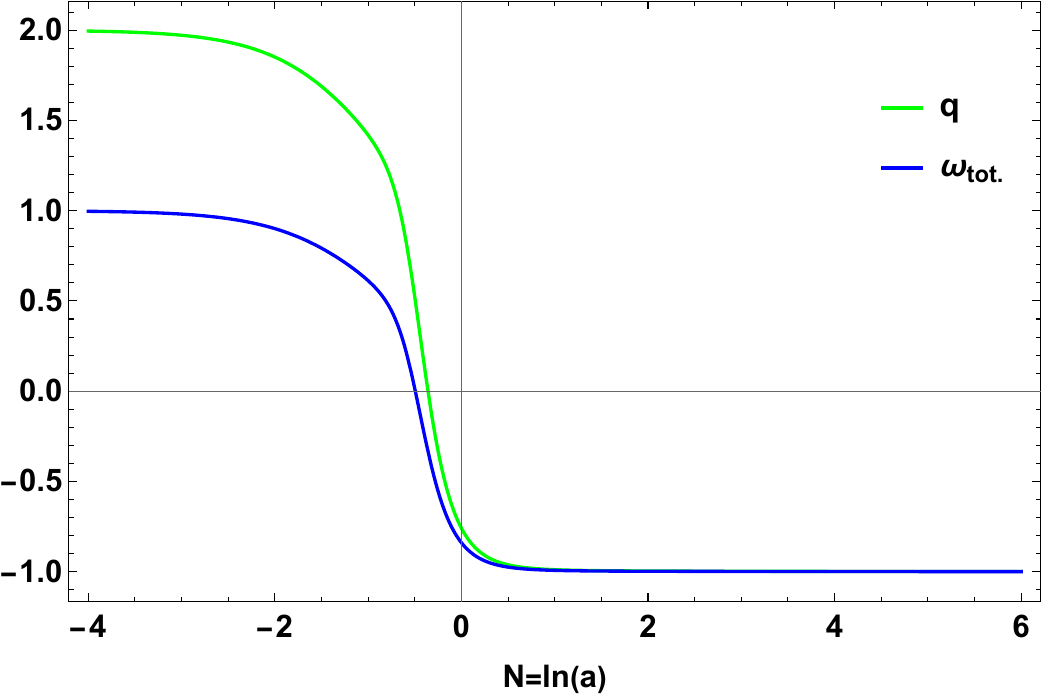}}
\caption{Evolutionary profile of the deceleration and the equation of state parameter for the case $n=2$ (left panel) and $n=-1$ (middle panel), with $\beta=1$, and for the case $n=-1$ with $\beta=\sqrt{3}$ (right panel) corresponding to the exponential potential.}\label{f3}
\end{figure}

\begin{table}[H]
\begin{center}\caption{Table shows the critical points and their behavior corresponding to the model $f(Q)=-Q+ \alpha Q^n$ with exponential potential $V(\phi)= V_0 e^{-\beta \phi}$.}
\begin{tabular}{|c|c|c|c|c|c|}
\hline
 Critical Points & Eigenvalues & Nature of critical point  & $q$ & $\omega$ \\
 $(x_c,y_c)$ &  $\lambda_1$ and $\lambda_2$ &  &  & \\
\hline 
$O(0,0)$ & $ -3, 0 $ & Stable  & $-1$ & $-1$ \\
$A(-1,0)$ & $6-6n, 3+\sqrt{\frac{3}{2}}\beta $ & Stable for $(n>1 \: \& \: \beta < -\sqrt{6} )$  & $2$ & $1$ \\
$B(1,0)$ & $6-6n, 3-\sqrt{\frac{3}{2}}\beta $ & Stable for $(n>1 \: \& \: \beta > \sqrt{6})$  & $2$ & $1$ \\
$C\left(\frac{\beta}{\sqrt{6}},\sqrt{1-\frac{\beta^2}{6}}\right)$ & $\frac{1}{2} (\beta^2-6), (1-n)\beta^2 $ & Stable for $(-\sqrt{6}< \beta < 0 \: \& \: n > 1 )$ or  $(0 < \beta < \sqrt{6} \: \& \: n > 1 )$  & $-1+\frac{\beta^2}{2}$ & $-1+\frac{\beta^2}{3}$ \\
$D\left(\frac{\beta}{\sqrt{6}},-\sqrt{1-\frac{\beta^2}{6}}\right)$ & $\frac{1}{2} (\beta^2-6), (1-n)\beta^2 $ & Stable for $(-\sqrt{6}< \beta < 0 \: \& \: n > 1 )$ or  $(0 < \beta < \sqrt{6} \: \& \: n > 1 )$  &$-1+\frac{\beta^2}{2}$ & $-1+\frac{\beta^2}{3}$ \\
\hline
\end{tabular}\label{Table-1}
\end{center}
\end{table}   
We present the asymptotic behavior of some prominent cosmological parameters such as the scale factor, the Hubble parameter, and the scalar field corresponding to obtained set of critical points,
\begin{itemize}
\item Corresponding to the case $x_c=0$ and $y_c=0$, we obtain the cosmic scalar factor as $a(t)=e^{H_0(t-t_0)}$, the Hubble parameter as $H(t)=H(t_0)=H_0$, and the scalar field $\phi(t)=\phi(t_0)=\phi_0$, where $t_0$ being the present time, and $a(t_0)=1$ and $H_0$ are being the present value of the scale factor and the Hubble parameter.
\item Corresponding to the case $x_c=\pm 1$ and $y_c=0$, we obtain the cosmic scalar factor as $a(t)=\left[ 3H_0(t-t_0)+1\right]^\frac{1}{3}$, the Hubble parameter as $H(t)=\frac{H_0}{3H_0(t-t_0)+1}$, and the scalar field $\phi(t)=\phi_0 \pm \frac{\sqrt{6}}{3} ln\left[ 3H_0(t-t_0)+1\right] $.
\item Corresponding to the case $x_c=\frac{\beta}{\sqrt{6}}$ and $y_c= \pm \sqrt{1-\frac{\beta^2}{6}}$, we obtain the cosmic scalar factor as $a(t)=\left[ \frac{\beta^2}{2}H_0(t-t_0)+1\right]^\frac{2}{\beta^2}$, the Hubble parameter as $H(t)=\frac{H_0}{\frac{\beta^2}{2}H_0(t-t_0)+1}$, and the scalar field $\phi(t)=\phi_0 + \left[ \frac{\beta^2}{2}H_0(t-t_0)+1\right]^\frac{2}{\beta} $.
\end{itemize}
The 2-D phase-space diagrams corresponding to autonomous system \eqref{s3}-\eqref{s4} for some specific parameter values are presented below in the Figure \eqref{f1}. In addition, the corresponding evolutionary profile of the scalar field and dark energy density are presented in the Figure \eqref{f2}. Further in the Figure \eqref{f3}, we present the corresponding evolutionary profile of the deceleration and the effective equation of state parameter.

Now we discuss the physical significance and stability of the critical points obtained for the specific parameter values as follows:
\begin{itemize}
\item \textbf{Case I ($n=2$ and $\beta=1$) :} In this case, the obtained set of critical points are $O_1(0,0)$, $A_1(-1,0)$, $B_1(1,0)$, $C_1\left( \frac{1}{\sqrt{6}}, \sqrt{\frac{5}{6}} \right)$, and $D_1\left( \frac{1}{\sqrt{6}}, -\sqrt{\frac{5}{6}} \right)$ with corresponding eigenvalues $(\lambda_1,\lambda_2)=(-3,0)$, $\left(-6,3+\sqrt{\frac{3}{2}}\right)$, $\left(-6,3-\sqrt{\frac{3}{2}}\right)$, $\left( -\frac{5}{2},-1 \right)$, and $\left( -\frac{5}{2},-1 \right)$ respectively. Moreover, the corresponding pair of values $(q,\omega)$ are $(-1,-1)$, $(2,1)$, $(2,1)$, $\left( -\frac{1}{2},-\frac{2}{3} \right)$, and $\left( -\frac{1}{2},-\frac{2}{3} \right)$. Therefore critical points $O_1$, $C_1$, and $D_1$ are stable, whereas points $A_1$ and $B_1$ are saddle. It should also note that the one can neglect the flow of trajectories below X-axis in the phase-space diagrams, since results are same as that of the above X-axis due to the symmetry in positive and negative axes of Y. Moreover, from the flow of phase-space trajectories in the Figure \eqref{f1} (left panel above), it is clear that we obtained a stable de-Sitter accelerated universe (represented by $O_1$) without any transition epoch. The same is reflected in the evolutionary profile of corresponding cosmological parameters presented in the Figures \eqref{f2} and \eqref{f3} (left panel).

\item \textbf{Case II ($n=-1$ and $\beta=1$) :} In this case, the obtained set of critical points are $O_2(0,0)$, $A_2(-1,0)$, $B_2(1,0)$, $C_2\left( \frac{1}{\sqrt{6}}, \sqrt{\frac{5}{6}} \right)$, and $D_2\left( \frac{1}{\sqrt{6}}, -\sqrt{\frac{5}{6}} \right)$ with corresponding eigenvalues $(\lambda_1,\lambda_2)=(-3,0)$, $\left(12,3+\sqrt{\frac{3}{2}}\right)$, $\left(12,3-\sqrt{\frac{3}{2}}\right)$, $\left( -\frac{5}{2},2\right)$, and $\left( -\frac{5}{2},2 \right)$ respectively. Moreover, the corresponding pair of values $(q,\omega)$ are $(-1,-1)$, $(2,1)$, $(2,1)$, $\left( -\frac{1}{2},-\frac{2}{3} \right)$, and $\left( -\frac{1}{2},-\frac{2}{3} \right)$. Therefore the critical point $O_2$ is stable, $A_2$ and $B_2$ are unstable, and points $C_2$ and $D_2$ are saddle. Further, from the flow of phase-space trajectories in the Figure \eqref{f1} (right panel above), we obtained evolutionary phase $A_2 \rightarrow O_2$ and $B_2 \rightarrow O_2$, both representing the evolution of the universe from stiff fluid dominated decelerated epoch to an accelerated de-Sitter epoch. The same is reflected in the evolutionary profile of corresponding cosmological parameters presented in the Figures \eqref{f2} and \eqref{f3} (middle panel). The results obtained for the parameter constraints in Case II is better than that of Case I.

\item \textbf{Case III ($n=-1$ and $\beta=\sqrt{3}$) :} In this case, the obtained set of critical points are $O_3(0,0)$, $A_3(-1,0)$, $B_3(1,0)$, $C_3\left( \frac{1}{\sqrt{2}}, \sqrt{\frac{1}{2}} \right)$, and $D_3\left( \frac{1}{\sqrt{2}}, -\sqrt{\frac{1}{2}} \right)$ with corresponding eigenvalues $(\lambda_1,\lambda_2)=(-3,0)$, $\left(12,3+\frac{3}{\sqrt{2}}\right)$, $\left(12,3-\frac{3}{\sqrt{2}}\right)$, $\left( -\frac{3}{2},6 \right)$, and $\left( -\frac{3}{2},6 \right)$ respectively. Moreover, the corresponding pair of values $(q,\omega)$ are $(-1,-1)$, $(2,1)$, $(2,1)$, $\left( \frac{1}{2},0 \right)$, and $\left( \frac{1}{2},0 \right)$. Therefore the critical point $O_3$ is stable, $A_3$ and $B_3$ are unstable, and points $C_3$ and $D_3$ are saddle. Further, from the flow of phase-space trajectories in the Figure \eqref{f1} (below), we obtained evolutionary phase $A_3 \rightarrow O_3$, $B_3 \rightarrow O_3$, and $A_3 \rightarrow C_3 \rightarrow O_3$. The evolutionary trajectory $A_3 \rightarrow C_3 \rightarrow O_3$ is quite interesting, representing the evolution of the universe from a decelerated stiff era to an accelerated de-Sitter era via matter dominated epoch. The same is reflected in the evolutionary profile of corresponding cosmological parameters presented in the Figures \eqref{f2} and \eqref{f3} (right panel). The results obtained for the parameter constraints in Case III is better among all three cases.
\end{itemize}

\subsection{Power-law potential}
\begin{figure}[h]
\includegraphics[scale=0.52]{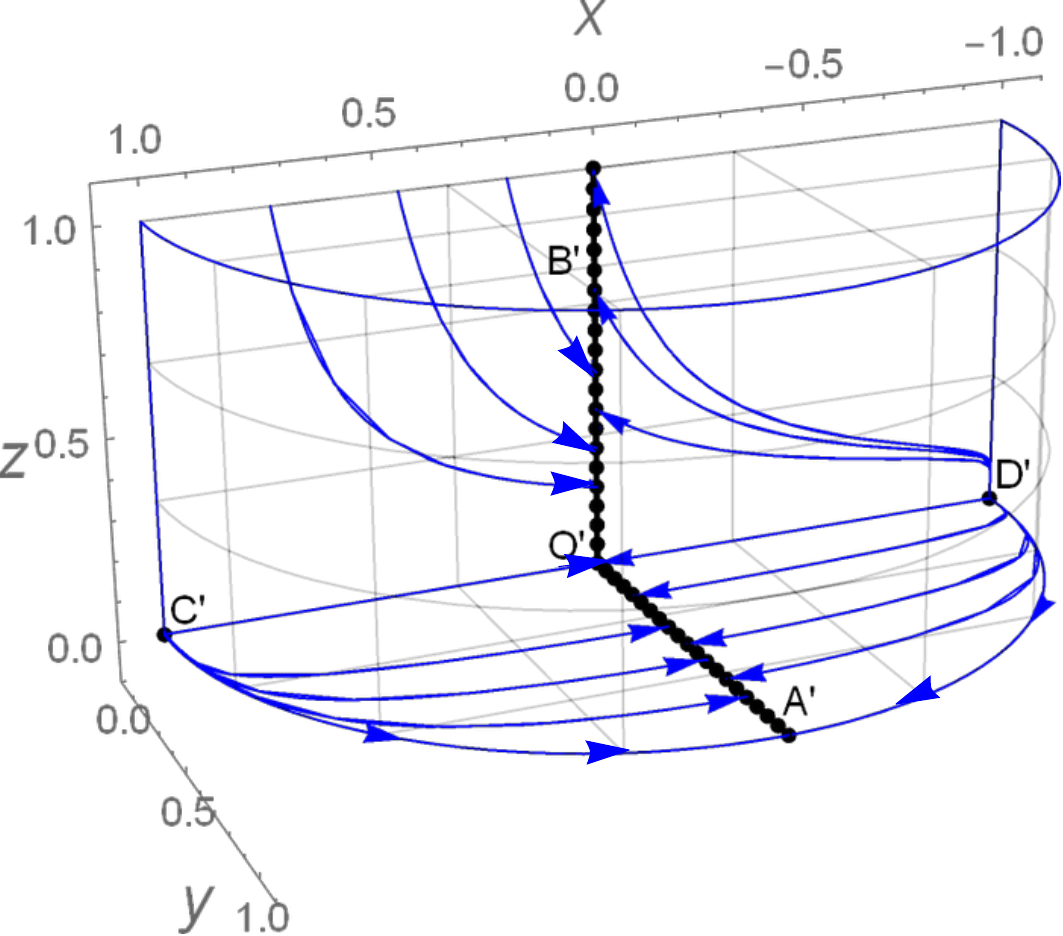}
\caption{The 3-D phase-space trajectories plotted for a set of solutions to the autonomous system for the parameter value  $n=-2$ and $k=0.16$ corresponding to the power-law potential.}\label{f4}
\end{figure}

\begin{figure}[H]
{\includegraphics[scale=0.479]{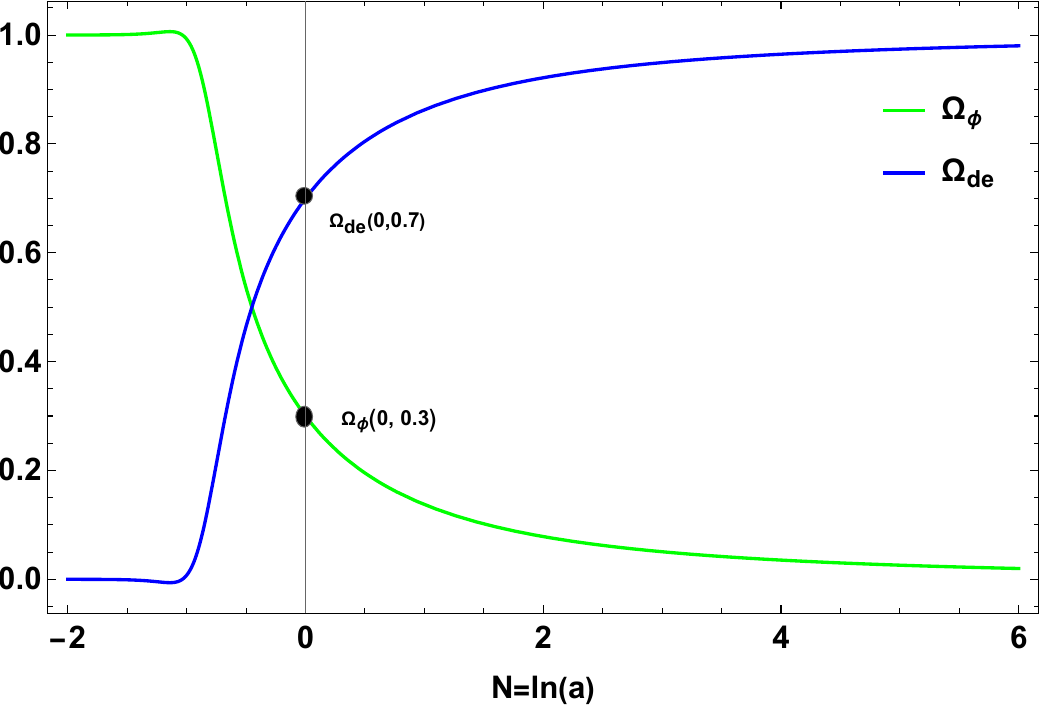}}
{\includegraphics[scale=0.49]{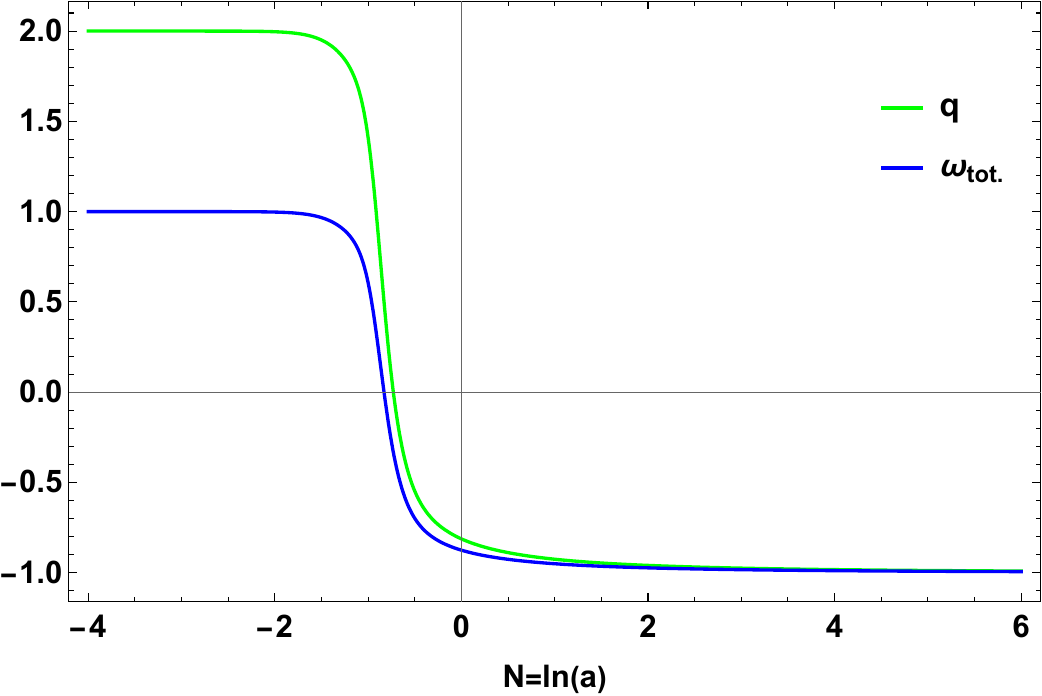}}
\caption{Evolutionary profile of scalar field density, dark energy density, deceleration, and the equation of state parameter for the parameter value $n=-2$ and $k=0.16$ corresponding to the power-law potential.}\label{f5}
\end{figure}

\justifying
We assume the following form of power-law potential,
\begin{equation}\label{r1}
V(\phi)= V_0\phi^{-k}
\end{equation}
The power-law type scalar field potentials have already been investigated in the context of both inflations \cite{BRT} and to address the cosmological constant problem \cite{PJE}. Such a potential naturally occurs in the string theory, such as law energy states of D-brains giving effective Tachyon field \cite{sen} and in Born-Infeld inspired lagrangian \cite{born}. Also, such potential has appeared in the tachyon-dominated backgrounds to probe the cosmological perturbations \cite{LRA}. Now, by using equation \eqref{r1}, the dynamical variable \eqref{4c} becomes 
\begin{equation}\label{r2}
\lambda=\frac{k}{\phi} \:\: \text{with} \:\: \Gamma= \frac{k+1}{k}
\end{equation}
,and hence the system \eqref{4d}-\eqref{4f} reduces to the following autonomous system,
\begin{equation}\label{r3}
x'=-3x \left[ 1+ \frac{x^2}{n \left( 1-x^2-y^2 \right)-1}  \right] + \sqrt{\frac{3}{2}} \lambda y^2
\end{equation}
\begin{equation}\label{r4}
y'=  \sqrt{\frac{3}{2}} \lambda xy \left[ 1- \frac{ \sqrt{6}x}{\lambda \{ n \left( 1-x^2-y^2 \right)-1 \} }  \right]
\end{equation}
\begin{equation}\label{r5}
\lambda'= - \frac{\sqrt{6}}{k} \lambda^2 x
\end{equation}
Note that the autonomous system \eqref{r3}-\eqref{r5} is invariant under the transformation $y \rightarrow -y$, and hence flow of trajectories obtained in the negative region of y would be a copy of  positive region. Moreover, the system \eqref{r3}-\eqref{r5} is also invariant under simultaneous transformations $x \rightarrow -x$ and $\lambda \rightarrow -\lambda$. Therefore, the physical region of the given dynamical system is represented by the positive-y half cylinder with infinite length from $\lambda=0$ to $\lambda= +\infty$. Hence, we define the following phase-space variable in order to compactify the variable $\lambda$,
\begin{equation}\label{r6}
z=\frac{\lambda}{\lambda+1}  \:\: \text{or} \:\: \lambda=\frac{z}{1-z}
\end{equation}
The new phase-space variable $z$ is bounded as $0 \leq z \leq 1$ and follows $z=0$ when $\lambda=0$ and $z=1$ when $\lambda \rightarrow + \infty$. Now, the autonomous system \eqref{r3}-\eqref{r5} reduces to,
\begin{equation}\label{r7}
x'=-3x \left[ 1+ \frac{x^2}{n \left( 1-x^2-y^2 \right)-1}  \right] + \sqrt{\frac{3}{2}} \frac{z}{(1-z)}y^2
\end{equation}
\begin{equation}\label{r8}
y'=  \sqrt{\frac{3}{2}} \frac{z}{(1-z)} xy \left[ 1- \frac{ \sqrt{6}x(1-z)}{z \{ n \left( 1-x^2-y^2 \right)-1 \} }  \right]
\end{equation}
\begin{equation}\label{r9}
z'= - \frac{\sqrt{6}}{k} z^2 x
\end{equation}
The critical points and their physical behavior, corresponds to autonomous system \eqref{r7}-\eqref{r9} are presented in the Table \eqref{Table-2}.
\begin{table}[H]
\begin{center}\caption{Table shows the critical points and their behavior corresponding to the model $f(Q)=-Q+ \alpha Q^n$ with potential $V(\phi)= V_0 \phi^{-k}$.}
\begin{tabular}{|c|c|c|c|c|c|}
\hline
 Critical Points $(x_c,y_c,z_c)$ & Eigenvalues ($\lambda_1$, $\lambda_2$,  $\lambda_3$) & Nature of critical point  & $q$ & $\omega$ \\
\hline 
$O'(0,0,0)$ & $ (-3,0,0) $ & Stable  & $-1$ & $-1$ \\
$A'(0,y,0)$ & $ (-3,0,0)  $ & Stable  & $-1$ & $-1$ \\
$B'(0,0,z)$ & $ (0,0,-3) $ & Stable  & $-1$ & $-1$ \\
$C'(1,0,0)$ & $ (3,0,6-6n) $ & Unstable for $ n \leq 1 $  and N.H. for $ n > 1 $   & $2$ & $1$ \\
$D'(-1,0,0)$ & $ (3,0,6-6n) $ & Unstable for $ n \leq 1 $  and N.H. for $ n > 1 $   & $2$ & $1$ \\
\hline
\end{tabular}\label{Table-2}
\end{center}
\end{table}   
We present the asymptotic behavior of some prominent cosmological parameters such as the scale factor, the Hubble parameter, and the scalar field corresponding to obtained set of critical points,
\begin{itemize}
\item Corresponding to the case $x_c=0$, we obtain the cosmic scalar factor as $a(t)=e^{H_0(t-t_0)}$, the Hubble parameter as $H(t)=H(t_0)=H_0$, and the scalar field $\phi(t)=\phi(t_0)=\phi_0$, where $t_0$ being the present time, and $a(t_0)=1$ and $H_0$ are being the present value of the scale factor and the Hubble parameter.
\item Corresponding to the case $x_c=\pm 1$ and $y_c=0$, we obtain the cosmic scalar factor as $a(t)=\left[ 3H_0(t-t_0)+1\right]^\frac{1}{3}$, the Hubble parameter as $H(t)=\frac{H_0}{3H_0(t-t_0)+1}$, and the scalar field $\phi(t)=\phi_0 \pm \frac{\sqrt{6}}{3} ln\left[ 3H_0(t-t_0)+1\right] $.
\end{itemize}
The 3-D phase-space diagram plotted for a set of solutions to the autonomous system \eqref{r7}-\eqref{r9} utilizing numerical approach for specific parameter value is presented in the Figure \eqref{f4}. Further, the corresponding evolutionary profile of the scalar field and dark energy density with the deceleration and the effective equation of state parameter are presented in the Figure \eqref{f5}. 

Now we discuss the physical significance and stability of the critical points obtained for the specific parameter value $n=-2$ and $k=0.16$. In this case, the obtained set of critical points are $O'(0,0,0)$, $A'(0,y,0)$ where $0 \leq y \leq 1$, $B'(0,0,z)$ where $0 \leq z \leq 1$, $C'(1,0,0)$, and $D'(-1,0,0)$ with corresponding eigenvalues $(\lambda_1,\lambda_2,\lambda_3)= (-3,0,0) $, $(-3,0,0) $, $(0,0,-3)$, $(3,0,18)$, and $(3,0,18)$ respectively. Moreover, the corresponding pair of values $(q,\omega)$ are $(-1,-1)$, $(-1,-1)$, $(-1,-1)$, $(2,1)$, and $(2,1)$. Therefore critical points $O'$, $A'$, and $B'$ are stable, whereas points $C'$ and $D'$ are unstable. Further, from the flow of trajectories in the 3-D phase-diagram presented in the Figure \eqref{f4}, we obtained evolutionary phase $C' \rightarrow O'$, $D' \rightarrow O'$, $C' \rightarrow A'$, $D' \rightarrow A'$, and $D' \rightarrow B'$. All the presented flow of trajectories exhibits identical behavior, and representing the evolution of the universe from a decelerated stiff era to an accelerated de-Sitter era. The same is reflected in the evolutionary profile of corresponding cosmological parameters presented in the Figure \eqref{f5}. The left panel of the Figure \eqref{f5} indicates that the scalar field density vanishes with the expansion of the universe, whereas the dark energy density dominates completely at the late-times. The right panel of the Figure \eqref{f5} indicates the transition from decelerated epoch to accelerated epoch of the universe in the recent past, and that end up with a de-Sitter universe at late-times. 

\section{Conclusion}\label{sec5}
\justifying

In this manuscript, we studied scalar field cosmology in the coincident $f(Q)$ gravity formalism. Modified gravity with non-metricity has been extensively investigated recently in different contexts such as late-time observational constraints, black holes, wormholes, and dynamical analysis, whereas the scalar field cosmology has been prominent in describing inflation as well as late-time acceleration. The modified $f(Q)$ function is now responsible for observed accelerating expansion, whereas the quantum field theory prediction of a tiny value of $\Lambda$ cannot be described by such a modified gravity scenario, and hence the addition of a scalar field take into account for quintessence scenario. In section \ref{sec2}, we presented the mathematical formulation of $f(Q)$ gravity with the Lagrangian density of the scalar field. Further, in section \ref{sec3}, we presented Friedmann-like equations of $f(Q)$ gravity ruling the gravitational interactions under the FLRW background in the presence of a scalar field. In section \ref{sec4}, we begin with a non-linear $f(Q)$ functional, specifically $f(Q)=-Q+\Psi(Q)=-Q+\alpha Q^n$, where $\alpha$ and $n$ are free model parameters. The assumed functional form is a polynomial correction to the STEGR case and has great significance in early and late-time cosmology. The considered $f(Q)$ function with $n > 1$ can potentially apply to the inflationary scenario, whereas the case $n < 1$ corrects the late-time cosmology. Further, we have defined phase-space variables (presented in equations \eqref{4a} and \eqref{4c}), and then we expressed the deceleration and the effective equation of state parameter in terms of phase-space variables. Now to probe the cosmological implications of the considered scenario, we assumed two specific forms of the potential function, specifically the exponential one $V(\phi)= V_0 e^{-\beta \phi}$ and the power-law $V(\phi)= V_0\phi^{-k}$, which are widely discussed in the literature. The corresponding critical points and their behaviors for the obtained autonomous systems are presented in Table \eqref{Table-1} and \eqref{Table-2}. Moreover, for both cases, the asymptotic behavior of the scale factor, the Hubble parameter, and the scalar field corresponding to the obtained set of critical points have been presented. In addition, we presented the 2-D phase-space diagrams corresponding to the exponential case for some parameter values, specifically, $n=2$ and $n=-1$ with $\beta=1$ , and $n=-1$ with $\beta=\sqrt{3}$, in the Figure \eqref{f1}. The behavior of corresponding cosmological parameters such as scalar field and dark energy density, deceleration, and the effective equation of state parameter is presented in Figure \eqref{f2} and \eqref{f3}. Then we have discussed the physical significance and stability of the obtained critical points corresponding to all three cases. We found that the results obtained for the parameter constraints in Case III are better among all three cases. We obtained evolutionary phase $A_3 \rightarrow C_3 \rightarrow O_3$ representing the universe's evolution from a decelerated stiff era to an accelerated de-Sitter era via matter-dominated epoch. Further, we have presented the 3-D phase-space diagram plotted for a set of solutions to the autonomous system corresponding to the power-law case for the parameter value $n=-2$ and $k=0.16$ (see Figure \eqref{f4}) and the behavior of corresponding cosmological parameters presented in the  Figure \eqref{f5}. In this case, all trajectories exhibit identical behavior, representing the universe's evolution from a decelerated stiff era to an accelerated de-Sitter era. Thus, we can conclude that the exponential case shows better evolution than the power-law case, and hence the present study successfully describes the different cosmological epochs of the universe. 

\section*{Data availability} There are no new data associated with this article.

\section*{Acknowledgments}

SG acknowledges Council of Scientific and Industrial Research (CSIR), Government of India, New Delhi, for junior research fellowship (File no.09/1026(13105)/2022-EMR-I). RS acknowledges University Grants Commission (UGC), New Delhi, India, for awarding a Senior Research Fellowship (UGC-Ref. No.: 191620096030). PKS  acknowledges the Science and Engineering Research Board, Department of Science and Technology, Government of India for financial support to carry out the Research project No.: CRG/2022/001847 and Transilvania University of Brasov for Transilvania Fellowship for Visiting Professors.

\end{document}